\documentclass[11pt]{article}

\usepackage{times}
\usepackage{natbib}
\bibpunct{(}{)}{;}{a}{,}{,}
\usepackage{fullpage}
\usepackage{graphicx}
\usepackage{amsmath,amssymb,amsthm}
\usepackage{bm}
\usepackage{latexsym}
\usepackage{hyphenat}
\usepackage[pdftex,colorlinks=true,linkcolor=blue,citecolor=blue,urlcolor=blue,bookmarks=false,pdfpagemode=None]{hyperref}
\usepackage{url}
\usepackage{verbatim}
\usepackage{fancyhdr}
\usepackage{setspace}
\usepackage{paralist}
\usepackage{boxedminipage}
\usepackage{rotating}

\newcommand{\bv}{\begin{array}}
\newcommand{\ev}{\end{array}}
\newcommand{\bit}{\begin{itemize}}
\newcommand{\eit}{\end{itemize}}
\newcommand{\ben}{\begin{enumerate}}
\newcommand{\een}{\end{enumerate}}
\newcommand{\beq}{\begin{equation}}
\newcommand{\eeq}{\end{equation}}
\newcommand{\bvq}{\begin{eqnarray}}
\newcommand{\evq}{\end{eqnarray}}
\newcommand{\ptoq}{p \rightarrow q}
\newcommand{\pfromq}{p \leftarrow q}

\begin{document}
\thispagestyle{empty}
\section*{\centerline{\Large Mixed membership analysis of high-throughput interaction studies}\vspace{20pt}\newline 
          \it \normalsize 
          Edoardo M. Airoldi, Princeton University~\footnote{Address correspondence to: Edo Airoldi, Carl Icahn Laboratory, Princeton University, Princeton, NJ 08544, USA.} \hfill (eairoldi@princeton.edu)\newline
          David M. Blei, Princeton University \hfill (blei@cs.princeton.edu)\newline
          Stephen E. Fienberg, Carnegie Mellon University \hfill (fienberg@stat.cmu.edu)\newline
          Eric P. Xing, Carnegie Mellon University \hfill (epxing@cs.cmu.edu)}

\begin{abstract}
 In this paper, we consider the statistical analysis of a protein interaction network. We propose a Bayesian model that uses a hierarchy of probabilistic assumptions about the way proteins interact with one another in order to: (i) identify the number of non-observable functional modules; (ii) estimate the degree of membership of proteins to modules; and (iii) estimate typical interaction patterns among the functional modules themselves.  Our model describes large amount of (relational) data using a relatively small set of parameters that we can reliably estimate with an efficient inference algorithm. We apply our methodology to data on protein-to-protein interactions in saccharomyces cerevisiae to reveal proteins' diverse functional roles.  The case study provides the basis for an overview of which scientific questions can be addressed using our methods, and for a discussion of technical issues.\newline

\noindent\textbf{Keywords:}  Data analysis; Bayesian inference; Latent Variables;
  Hierarchical mixture model; Variational Expectation-Maximization;
  Mean-field approximation; Relational data; Unipartite graphs. 
\end{abstract}

%%
%  cite wong on biometrics, and bonnie berger on psb 2007 (sontag, singh)
%%

\section{Introduction}

Relational data, which describe measurements on pairs of objects, arise in a variety of applications. Citation networks underlying scientific collections of papers are obtained from references, which connect pairs of papers; web-graphs are obtained from hyperlinks, which connect pairs of web-pages; protein networks are obtained from physical interaction records, which relate pairs of proteins.  In this paper, the discussion develops intuitions for protein interaction networks obtained experimentally with yeast two-hybrid tests and others means.

There are important differences between models for relational data,
such as protein-protein interactions, and non-relational data, such as
protein attributes.  Specifically, the exchangeability assumptions
underlying models of non-relational data are typically violated by
relational data \citep{Airo:Blei:Fien:Xing:2006b}.  Descriptive data analyses of
relational measurements consider a rich set of goals, which often
include: (i) the identification of the number of non-observable groups
of objects, e.g., functional modules or stable protein complexes; (ii) the estimation of the
degree of membership of objects to groups, e.g., of protein to functional modules or protein
complexes; and (iii) the estimation of typical interaction patterns
among the groups themselves, e.g., among functional modules or protein complexes. While the first two tasks arise in non-relational data settings as well, the last task is specific to relational data settings.
 In addition to these descriptive goals we are often interested in inferring latent quantities that are useful for making predictions. In the context of protein interaction networks, we want to identify group memberships and interaction patterns that are instrumental in predicting new relations and object specific attributes; more specifically, one may try to predict interactions between pairs of proteins and individual proteins' functional annotations, using
patterns of interaction between them, and between the stable protein
complexes they belong to.

\subsection{Novel Contributions}

% !!! DMB: add a mixed membership model citation (erosheva) somewhere
% here

 In this paper, we propose the {\it Admixture of Latent Blocks} (ALB),
where proteins exhibit membership in multiple latent groups.  We
develop efficient posterior inference algorithms for discovering the
membership of proteins to groups from large collections of observed
protein-protein interaction data.

 In the context of protein interaction networks, {\it mixed membership}
relaxes the mixture modeling assumption that each protein belongs to a
single group.  ALB uses the mixed-membership of the proteins to
explain interactions measured between them.  Specifically, a latent
{\it stochastic block structure} allows us to model interaction
patterns among the groups by encoding the probabilities according to
which pairs of individual proteins interact as generic members of the
corresponding pairs of groups.

 We develop an efficient inference algorithm based on variational methods. This provides a fast alternative to MCMC, and allows us to analyze the large collections of relational data that arise in biological applications.
 
% In the context of variational methods, we develop a novel inference
% algorithm, which we call {\it nested variational}. This algorithm is
% better suited for carrying out the inference in Bayesian hierarchical
% models of relational data when compared to the na\"ive variational
% algorithm (\cite*{Jord:Ghah:Jaak:Saul:1999}).  The na\"ive variational
% algorithm has been applied to non-relational data in the published
% literature; hence we make a note of a few reasons that can potentially
% lead to a poor approximation~\footnote{The na\"ive variational
%   algorithm converges, but entails a poor approximation of the
%   posterior of non-observable quantities of interest, e.g., the random
%   vectors that encode the mixed membership of proteins to protein
%   complexes.} when it is used in relational data settings.

 We apply our methodology to a large protein interaction network to
reveal proteins' diverse functional roles.  The case study in Section
\ref{sec:protein_data} illustrates the scientific questions that can be
addressed with our model, the alternative inference and estimations
strategies, and the technical issues that arise in such analyses.

\subsection{Related Work}

 There is a history of probabilistic models for relational data analysis in Statistics.  Part of this literature is rooted in the stochastic block modeling ideas from psychometrics and sociology. This model is due to \cite{Holl:Lein:1975}, and was later elaborated upon by others \citep[see, e.g.,][]{Fien:Meye:Wass:1985,Wass:Patt:1996,Snij:2002}. In machine learning, Markov random networks have been used for link prediction \citep{Task:Wong:Abbe:Koll:2003} and the traditional block models from Statistics have been extended with nonparametric Bayesian priors \citep{Kemp:Grif:Tene:2004,Kemp:Tene:Grif:etal:2006}.

Mixed membership models for clustering have emerged as a powerful and
popular analytical tool for analyzing large databases involving text
\citep{Hofm:1999,Blei:Jord:Ng:2003}, text and references
\citep{Cohn:Hofm:2001,Eros:Fien:Laff:2004}, text and images
\citep{Barn;Duyg;Frei;Fors;2003}, multiple disability measures
\citep{Eros:Fien:2005,Mant:Wood:Toll:1994}, and genetics information
\citep{Rose:Prit:Webe:etal:2002,Prit:Step:Donn:2000,Xing:Ng:Jord:Russ:2003}. These models use a simple generative
model, such as bag-of-words or naive Bayes, embedded in a larger
hierarchical model that involves a latent variable structure.  This
induces dependencies between the observed data, and introduces
statistical control over the estimation of what might otherwise be an
extremely large set of parameters.

\section{The Scientific Problem}

Our goal is to analyze proteins' diverse functional roles by analyzing
their local and global patterns of interaction.  The biochemical
composition of individual proteins make them suitable for carrying out
a specific set of cellular operations, or \textit{functions}.
Proteins typically carry out these functions as part of stable protein
complexes \citep{Krog:Cagn:Yu:Zhon:2006}.  There are many
situations in which proteins are believed to interact
\citep{Albe:John:Lewi:Raff:2002}; the main intuition behind our
methodology is that pairs of protein interact because they are part of
the same stable protein complex, i.e., co-location, or because they
are part of interacting protein complexes as they carry out compatible
cellular operations.

\subsection{Protein Interactions and Functional Annotations}

 The Munich Institute for Protein Sequencing (MIPS) database was
created in 1998 based on evidence derived from a variety of
experimental techniques, but does not include information from
high-throughput data sets \citep{Mewe:Amid:Arno:etal:2004}.
 It contains about 8000 protein complex associations in yeast.  We analyze
a subset of this collection containing 871 proteins, the interactions
amongst which were hand-curated. The institute also provides a set of
functional annotations, alternative to the gene ontology (GO).  These
annotations are organized in a tree, with 15 general functions at the
first level, 72 more specific functions at an intermediate level, and
255 annotations at the the leaf level.  In Table \ref{tab:functions}
we map the 871 proteins in our collections to the main functions of
the MIPS annotation tree; proteins in our sub-collection have about
$2.4$ functional annotations on average.\footnote{We note that the
  relative importance of functional categories in our sub-collection,
  in terms of the number of proteins involved, is different from the
  relative importance of functional categories over the entire MIPS
  collection.}

\begin{table}[ht!]
\begin{center}
\begin{tabular}{rlr|rlr}
\# & Category                           & Count & \# & Category                           & Count \\ \hline
 1 & Metabolism                         &   125 &  9 & Interaction w/ cell. environment   &    18 \\
 2 & Energy                             &    56 & 10 & Cellular regulation                &    37 \\
 3 & Cell cycle \& DNA processing       &   162 & 11 & Cellular other                     &    78 \\
 4 & Transcription (tRNA)               &   258 & 12 & Control of cell organization       &    36 \\
 5 & Protein synthesis                  &   220 & 13 & Sub-cellular activities            &   789 \\
 6 & Protein fate                       &   170 & 14 & Protein regulators                 &     1 \\
 7 & Cellular transportation            &   122 & 15 & Transport facilitation             &    41 \\
 8 & Cell rescue, defence \& virulence  &     6 &    &                                    &         \\
\end{tabular}
\end{center}
\caption{The 15 high-level functional categories obtained by cutting the MIPS annotation tree at the first level and how many proteins (among the 871 we consider) participate in each of them. Most proteins participate in more than one functional category, with an average of $\approx 2.4$ functional annotations.}
\label{tab:functions}
\end{table}

\begin{figure}[t!]
  \centering
   \includegraphics[width=15cm]{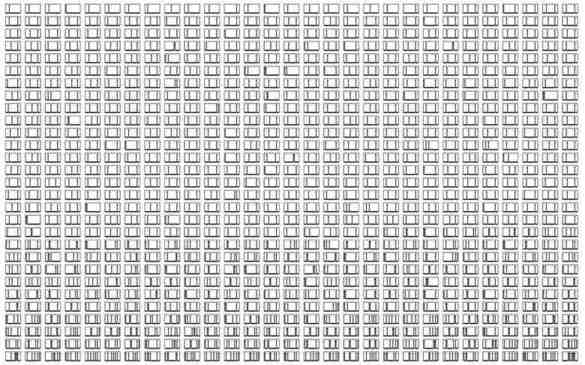}
  \centering
   \includegraphics[width=15cm]{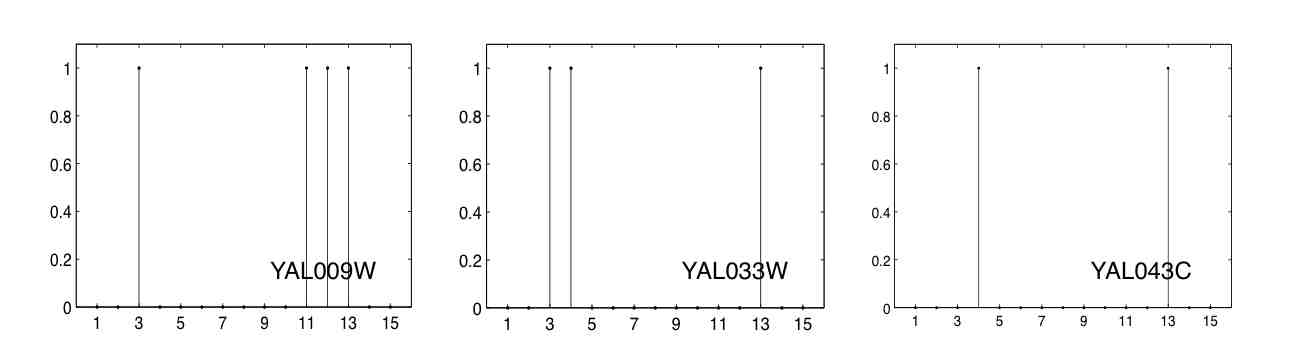}
  \caption{By mapping individual proteins to the 15 general
    functions in Table \ref{tab:functions}, we obtain a 15-dimensional representation for each
    protein.  Here, each panel corresponds to a protein; the
    15 functional categories are displayed on the $X$ axis, whereas
    the presence or absence of the corresponding functional annotation
    is displayed on the $Y$ axis.  The plots at the bottom zoom into three example panels (proteins).}
\label{fig:mipsdata}
\end{figure}

By mapping proteins to the 15 general functions, we obtain a
15-dimensional representation for each protein.  In Figure
\ref{fig:mipsdata} each panel corresponds to a protein; the 15
functional categories are ordered as in Table \ref{tab:functions} on
the $X$ axis, whereas the presence or absence of the corresponding
functional annotation is displayed on the $Y$ axis.

\section{The Admixture of Latent Blocks Model}
\label{sec:model}

The admixture of latent blocks (ALB) models observed protein
interaction networks, $G_{1:M} = ( R_{1:M}, \mathcal{P} )$.  The
presence or absence of a physical interaction among pairs of proteins
$p,q \in \mathcal{P}$ is measured over $M$ distinct experimental
conditions and encoded by Bernoulli random variables $R_m (p,q)$.  Let
us assume that the we observe networks among $N:=|\mathcal{P}|$
distinct proteins.

\subsection{Mixed Membership}

Stable protein complexes are believed to play a major role in
cellular processes \citep{Krog:Cagn:Yu:Zhon:2006}.  As a
consequence, protein interaction networks provide insights into
individual protein's functionality to the extent to which they carry
information about the membership of individual proteins to stable protein complexes.  In
a complex biological system, many proteins are functionally versatile
and distinct copies of the same protein participate in multiple
protein complexes, which perform different cellular processes (i.e.,
functions) at different times or under different biological
conditions.  Thus, when modeling interactions as observable outcomes
of latent functional processes, it is natural to adopt a flexible
model which allows distinct copies of a protein to interact with other
proteins in multiple, functionally related biological contexts.

For example, a signal transduction protein may sometimes interact with
a cellular membrane protein as part of a signal receptor; at another
time, it may interact with the transcription complex as an auxiliary
transcription factor.  Furthermore, there is direct empirical evidence
that individual proteins may perform multiple functions while taking
part in a cellular process,~\footnote{As an example, see the data
  about functional annotations of about 850 proteins in Yeast is
  presented in Figure \ref{fig:mipsdata}.  The data was provided,
  courtesy the Munich Institute for Protein Sequencing
  (\cite*{Mewe:Amid:Arno:etal:2004}).} and that they typically carry
them out as members of stable protein complexes.

The mixed membership assumption provides our model with such a
desirable feature.  Under this assumption, we introduce mixed
membership vectors $\vec{\pi}_{1:N}$, such that $\pi_{nk}$ is the
probability according to which copies of the $n$-$th$ protein
participate into copies of the $k$-$th$ protein complex, and $\sum_k \pi_{nk}
=1$.  In the ALB model, mixed membership is a global feature of the behavior
of a protein, i.e., it emerges from the composition of the collection
of individual interactions a protein is involved with.  Each these
individual interactions is characterized by a {\it single membership}
of each of the two proteins involved to a pair of stable protein
complexes.

Such single memberships of pairs of proteins to pairs of stable
protein complexes for one observed interaction, $R_m(p,q)$, are
encoded in one corresponding pair of latent protein complex
indicators, $(\vec{z}_{\ptoq}^m,\vec{z}_{\pfromq}^m)$.  These
interaction-specific indicators induce flexibility in the model both at the protein level and at the interaction level. Specifically, (i) distinct copies of the same
protein to interact with other proteins as a member of different
stable protein complexes, e.g., under the same experimental
conditions; and (ii) distinct interactions between copies of the same
pair of proteins to be the expression of interactions of different
pairs of stable protein complexes, e.g., under different experimental
conditions.  In the ALB model, each latent indicator $\vec{z}_{\ptoq}^m$
$($respectively, $\vec{z}_{\pfromq}^m)$ is a multinomial random vector
with unitary size parameter, so that a single membership is sampled
 for each unit measurement, and with probabilities $\vec{\pi}_{p}$
$($respectively, $\vec{\pi}_q)$ to constrain such single memberships
to follow the appropriate mixed membership profile of the $p$-$th$ protein to the $K$ protein complexes.

 During estimation and inference, the model recovers the non-observable stable protein complexes that are likely to carry out the functional processes
underlying the data, in terms of the degree to which proteins in
$\mathcal{P}$ take part in them, i.e., in terms of the mixed
membership vectors $\vec{\pi}_{1:N}$, by assessing the similarity of observed protein-to-protein interaction patterns.

\subsection{Latent Blocks}

In the experimental setting where measurements of physical
interactions are taken, a certain number of stable protein complexes
exist.  The number of functions underlying the data, which are carried
out by these complexes, as well as the interaction patterns among such
complexes are also of primary interest.

 A scalar parameter, $K$, encodes the number of non-observable protein
functions underlying the collection of observed interactions in the ALB model. Assuming that $K$ distinct functions exist, a
latent block structure $B$ encodes the interaction patterns among the
$K$ distinct stable protein complexes which carry them out.  The
latent block structure is a table of size $(K \times K)$.  The generic
entry $B(g,h)$ in the table encodes the probability according to which
the pair $(p,q)$ of proteins interacts, whenever a copy of the
$p$-$th$ protein is a member of the $g$-$th$ complex and a copy of
the $q$-$th$ protein is a member of the $h$-$th$ complex.

 During estimation and inference, the model is fitted assuming that a pre-specified number of functions
$K$ underlies the data, and recovers the probabilities according to which
pairs of individual proteins interact as generic members of the
corresponding pairs of stable protein complexes, i.e., the interaction
patterns among stable protein complexes $B$.

\subsection{The Data Generating Process}
\label{sec:dgp}

 The data generating process for $M$ protein interaction networks, $G_{1:M}$, assumes that the number of
individual proteins, $N:=|\mathcal{P}|$, the number of protein
complexes, $K$, their interaction patterns, $B$, and the average mixed
membership of protein to functions, $\vec{\alpha}$, are given
a-priori.

\begin{figure}[t!]
  \centering
  \includegraphics[width=\textwidth]{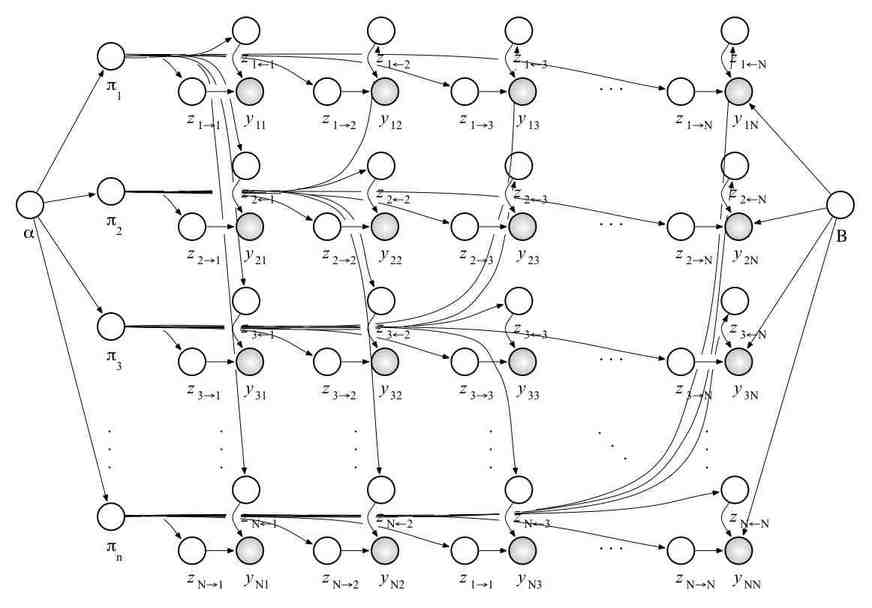}
  \caption{A graphical representation of the admixture of latent blocks (ALB) model. The boxes are \textit{plates} and represent replicates of observed networks. White nodes represent latent variables, whereas shaded (gray) nodes represent observed interactions.}
\label{fig:graphical_model}
\end{figure}

 The process then posits that, within the $m$-$th$ network, each observed interaction $R_m(p,q)$, $p,q \in \mathcal{P}$, is a Bernoulli random variable ~\footnote{Note that the data generating process does not necessarily generate  symmetric interactions, $R_m(p,q) = R_m(q,p)$, thus increasing the applicability of our methodology to relational data that arise in other domains.} with probability of success $\sigma_{pq}^m$.  The \textit{single memberships}, $(\vec{z}_{\ptoq}^{m}, \vec{z}_{\pfromq}^m)$, and the latent block structure, $B$, provide
competing explanations of the (scalar) probability of success
of each observed interaction, $R_m(p,q)$, which is defined as follows,
\[
 \sigma_{pq}^m = \vec{z}_{\ptoq}^{m~\top} B ~ \vec{z}_{\pfromq}^m.
\]
  The collection of single
membership indicators of a protein $p$, $\{ \vec{z}_{\pfromq}^m : q
\in \mathcal{P}, m=1,\dots,M \}$, each a multinomial random vector
with parameters $(\vec{\pi}_p,1)$, is constrained by the global mixed membership
behavior encoded in $\vec{\pi}_p$.  The \textit{mixed memberships} and
the latent block structure provide {\it competing} explanations of the
average probability of success of an interaction, given by $\bar
\sigma_{pq} = \vec{\pi}_p^{~\top} B ~ \vec{\pi}_q$, across the $M$
experimental conditions.  
 The mixed membership vectors are further constrained by positing they are (non-observable, independent and identically distributed) samples from a common Dirichlet distribution with hyper-parameter $\vec{\alpha}$.

 The data generating process for $G_{1:M} = ( R_{1:M}, \mathcal{P} ) $
maps a small set of constants to the data, $DGP : (\mathcal{P},M,K,\vec{\alpha},B) \rightarrow R_{1:M}$, and it is instantiated as follows.
\begin{enumerate}
\item For each protein $p \in \mathcal{P}$
 \begin{enumerate}
  \item[1.1.] Sample $\vec{\pi}_p \sim Dirichlet~(\vec{\alpha})$.
 \end{enumerate}
 \item For each protein interaction network $m = 1, \dots, M$
 \begin{enumerate}
  \item[2.1.] For each pair of proteins $(p,q) \in \mathcal{P} \otimes \mathcal{P}$
  \begin{enumerate}
   \item[2.1.1.] Sample group $\vec{z}_{\ptoq}^m \sim Multinomial~( \vec{\pi}_p , 1 )$
   \item[2.1.2.] Sample group $\vec{z}_{\pfromq}^m \sim Multinomial~( \vec{\pi}_q , 1 )$
   \item[2.1.3.] Sample $R_m(p,q) \sim Bernoulli~( \vec{z}_{\ptoq}^{m~\top} B ~ \vec{z}_{\pfromq}^m )$
  \end{enumerate}
 \end{enumerate}
\end{enumerate}

 The process above suggests a hierarchical decomposition of the joint
probability distribution of the observations, $R_{1:M}$, and the
latent variables,~\footnote{Two sets of latent protein complex indicators corresponding to the $m$-$th$ protein network are denoted compactly by, $\{ \vec{z}_{\ptoq}^m ~:~ p,q \in \mathcal{P} \} =: Z^\rightarrow_{m}$ and $\{ \vec{z}_{\pfromq}^m ~:~ p,q \in \mathcal{P} \} =: Z^\leftarrow_{m}$.}
$(\vec{\pi}_{1:N},Z^\rightarrow_{1:M},Z^\leftarrow_{1:M})$; that is, the integrand in Equation \ref{eq:likelihood_with_zs}.
 By integrating the latent variables out of the joint we obtain the likelihood of
the observations,
\begin{eqnarray}
\label{eq:likelihood_with_zs}
 p ( R_{1:M} | \vec{\alpha},B ) & = & \int_{\Pi \otimes \mathcal{Z}} ~ p ( R_{1:M},\vec{\pi}_{1:N},Z^\rightarrow_{1:M},Z^\leftarrow_{1:M} | \vec{\alpha},B ) ~ d \vec{\pi} ~ d Z \\
                                & = & \int_{\Pi \otimes \mathcal{Z}} \biggm( \prod_m \prod_{p,q} ~ p_1 ( R_m(p,q) | \vec{z}_{\ptoq}^m, \vec{z}_{\pfromq}^m, B)  ~ p_2 (\vec{z}_{\ptoq}^m | \vec{\pi}_p,1) \times \nonumber \\
                                &   & \times ~ p_2 (\vec{z}_{\pfromq}^m | \vec{\pi}_q,1) \biggm) \prod_p ~ p_3 (\vec{\pi}_p | \vec{\alpha}) ~ d \vec{\pi} ~ d Z \nonumber
\end{eqnarray}
where $p_1$ is Bernoulli, $p_2$ is multinomial, and $p_3$ is
Dirichlet.  A graphical representation using plates of the statistical
models corresponding to data generating process is given in Figure
\ref{fig:graphical_model}.

  A recurring question, which bears relevance to mixed membership models in general, is why one does not necessarily want to integrate out the single membership indicators---$(\vec{z}_{\ptoq}^{m}, \vec{z}_{\pfromq}^m)$ in the specifications above.
 There are some computational aspects to this but a practical issue that argues against such marginalization is that we would often lose interpretable quantities that are useful for making predictions, for de-noising new measurements, or for performing other tasks. In fact, the posterior distributions of such quantities typically carry substantive information about elements of the application at hand. In the application to protein interaction networks, for example, they encode the interaction-specific memberships of individual proteins to protein complexes.

\subsection{Modeling Rare Interactions}
\label{sec:rare_interactions}

The specifications of the data generating process suggests that observations about
interactions and non-interactions are equally important in terms of
their contributions to model fitness, e.g., see the integrand in
Equation \ref{eq:likelihood_with_zs}.  In other words, they equally
compete for a likely explanation in terms of estimates for
$(\vec{\alpha},B,\vec{\pi}_{1:N})$.  As a consequence, in experimental
settings where interactions are rare, the estimation and inference
tasks will find hyper-parameter values and posterior distributions
that explain patterns of non-interaction rather than patterns of
interaction.

In order to be able to calibrate the importance of rare interactions,
we introduce the sparsity parameter $\rho \in [0,1]$, which models how
often a non-interaction is due to noise and how often it carries
information about proteins' memberships to protein complexes.  This
leads to a new generative process, where the non-interactions are
generated from a mixture between the original Bernoulli (step 2.1.3.\
in the data generating process) and a point mass at zero, with weights $(1-\rho)$
and $\rho$ respectively.  The probabilities of non-interactions are
set to,
\[
1-\sigma_{pq}^m = (1-\rho) \cdot \vec{z}_{\ptoq}^{m~\top} (1-B) ~
\vec{z}_{\pfromq}^m + \rho,
\]
and the probabilities of successful interactions in the data generating process become,
\begin{eqnarray*}
  \sigma_{pq}^m & = & (1-\rho) \cdot \vec{z}_{\ptoq}^{m~\top} B ~ \vec{z}_{\pfromq}^m \\
  & = & \vec{z}_{\ptoq}^{m~\top} B' ~ \vec{z}_{\pfromq}^m,
\end{eqnarray*}
where $B'(g,h) = (1-\rho) \cdot B(g,h)$ for $g,h=1,\dots,K$.

 During estimation and inference, a large value of $\rho$ leads to interactions in the matrix being weighted more than non-interactions to the extent of informing the estimates of $(\vec{\alpha},B,\vec{\pi}_{1:N})$. In fact, when $\rho \approx 1$ the most likely explanation for non-interactions is generic noise, i.e., zeros are likely to be generated from the point mass.

\subsection{Parameter Estimation and Posterior Inference}
\label{sec:infest}

 We develop estimation strategies for the hyper-parameters $(\vec{\alpha},B)$ within the empirical Bayes framework \citep{Morr:1983,Carl:Loui:2005}. We develop a variational approximation to Expectation-Maximization (EM) to carry out posterior inference for the latent mixed-membership vectors, $\vec{\pi}_{1:N}$. A description of the algorithm and the mathematical derivations are presented elsewhere \citep{Airo:Blei:Fien:Xing:2006b}. The optimal number of blocks, $K$, is selected via cross-validation on the held-out likelihood.

 Briefly, in order to estimate $(\vec{\alpha},B)$ and infer posterior
distributions for $\vec{\pi}_{1:N}$ we need to be able to evaluate the
likelihood, which involves the non-tractable integral in Equation
\ref{eq:likelihood_with_zs}.  Given the large amount of data available
about biological (e.g., protein) networks, approximate posterior
inference strategies are considered in the context of variational
methods; a computationally cheaper alternative to Monte Carlo Markov
chain methods.  Using variational methods, we find a tractable lower
bound for the likelihood that can be used as a surrogate for inference
purposes.  This leads to approximate MLEs for the hyper-parameters and
approximate posterior distributions for the (latent) mixed-membership
vectors.

We introduce a variant of variational EM for our model, which we term
\textit{nested variational} EM algorithm.  Our algorithm improves the
na\"ive variational EM in two aspects: (i) it is parallelizable when
applied to relational data; and (ii) it reduces memory requirements
from $(NK+N^2K)$ to $(NK+K)$ per iteration.  

\section{Application to Protein Interactions in \emph{Saccharomyces Cerevisiae}}
\label{sec:protein_data}

Protein-protein interactions (PPI) form the physical basis for the
formation of complexes and pathways that carry out different
biological processes.  A number of high-throughput experimental
approaches have been applied to determine the set of interacting
proteins on a proteome-wide scale in yeast. These include the
two-hybrid (Y2H) screens and mass spectrometry methods.  Mass spectrometry can be used to identify components of protein complexes
\citep{Gavi:Bosc:Krau:etal:2002,Ho:Gruh:Heil:etal:2002}.

High-throughput methods, though, may miss complexes that are not
present under the given conditions.  For example, tagging may disturb
complex formation and weakly associated components may dissociate and
escape detection.  Statistical models that encode information about
functional processes with high precision are an essential tool for
carrying out probabilistic de-noising of biological signals from
high-throughput experiments.

Our goal is to identify the proteins' diverse functional roles by
analyzing their local and global patterns of interaction via ALB.  The
biochemical composition of individual proteins make them suitable for
carrying out a specific set of cellular operations, or
\textit{functions}.  Proteins typically carry out these functions as
part of stable protein complexes \citep{Krog:Cagn:Yu:Zhon:2006}.
There are many situations in which proteins are believed to interact
\citep{Albe:John:Lewi:Raff:2002}. The main intuition behind our
methodology is that pairs of protein interact because they are part of
the same stable protein complex, i.e., co-location, or because they
are part of interacting protein complexes as they carry out compatible
cellular operations.

\subsection{Brief Summary of Previous Findings}
\label{sec:previously}

In previous work, we established the usefulness of an admixture of
latent blockmodels for analyzing protein-protein interaction data
\citep{Airo:Blei:Xing:Fien:2005}. For example, we used the ALB for
testing functional interaction hypotheses (by setting a null
hypothesis for $B$), and unsupervised estimation experiments. In the next Section, we assess whether, and how much, functionally relevant biological signal can be captured in by the ALB.

\begin{figure}[b!]
  \centering
   \includegraphics[width=7.5cm]{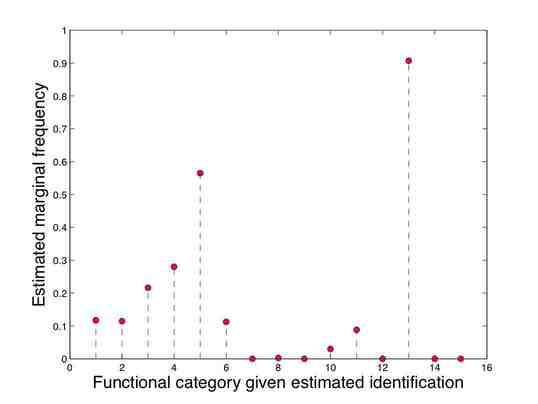}
   \includegraphics[width=7.5cm]{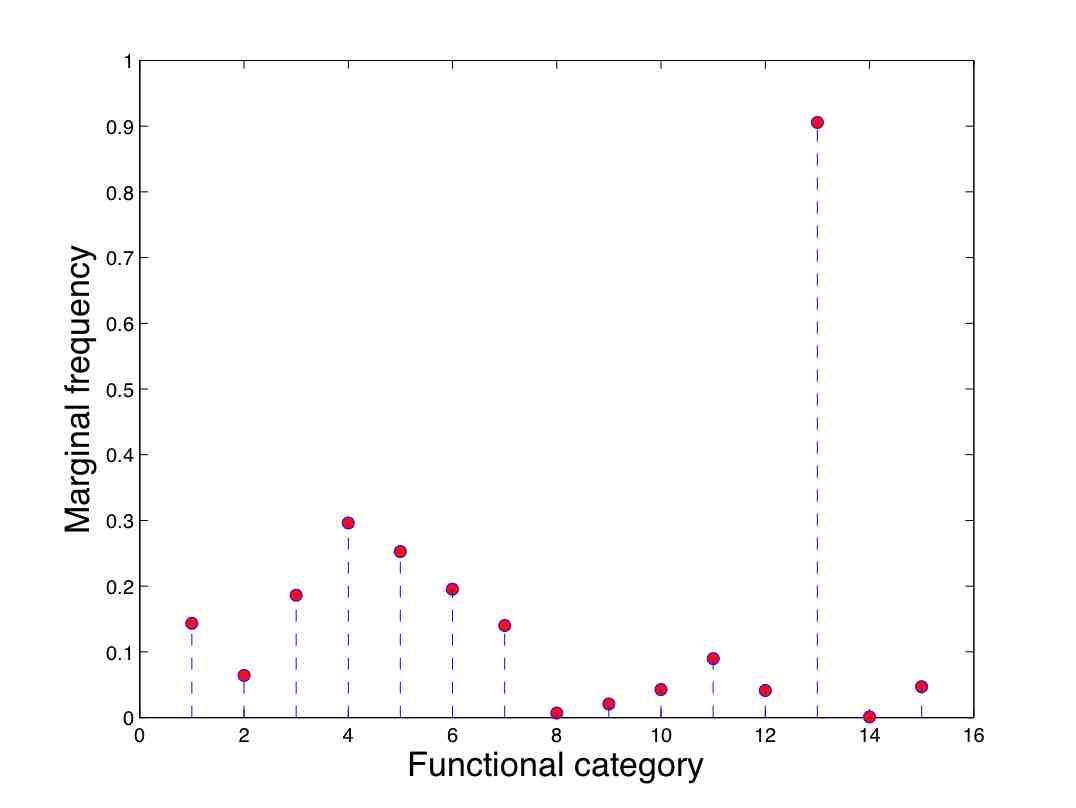}
  \caption{We estimate the mapping of latent groups to functions.  The
    two plots show the marginal frequencies of membership of proteins
    to true functions (bottom) and to identified functions (top), in
    the cross-validation experiment.  The mapping is selected to
    maximize the accuracy of the predictions on the training set, in
    the cross-validation experiment, and to minimize the divergence
    between marginal true and predicted frequencies if no training
    data is available---see Section \ref{sec:previously}.}
\label{fig:identify}
\end{figure}
\begin{figure}[t!]
  \centering
   \includegraphics[width=14cm]{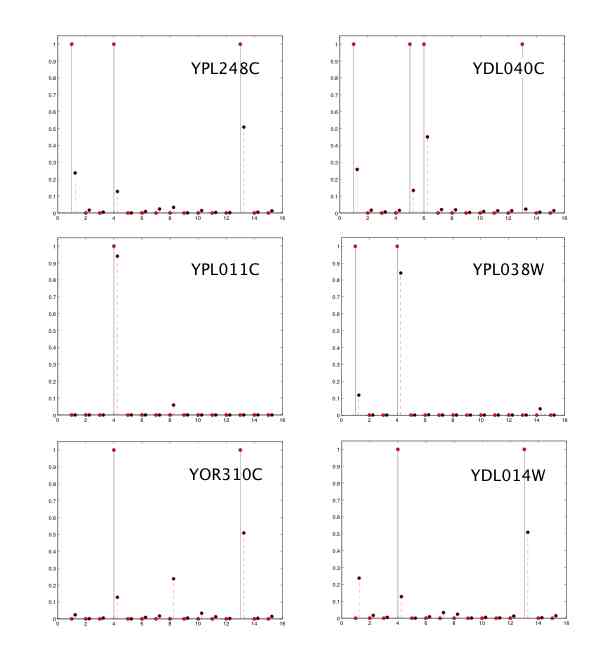}
 \caption{Predicted mixed-membership probabilities (dashed, red lines)
   versus binary manually curated functional annotations (solid, black
   lines) for 6 example proteins.  The identification of latent groups
   to functions is estimated, Figure
   \ref{fig:identify}.}
\label{fig:exampleprot}
\end{figure}

In summary, the results in \citet{Airo:Blei:Xing:Fien:2005} show that the ALB identifies protein complexes
whose member proteins are tightly interacting with one another.  The
identifiable protein complexes correlate with the following four
categories of Table \ref{tab:functions}: cell cycle \& DNA processing,
transcription, protein synthesis, and sub-cellular activities.  The
high correlation of inferred protein complexes can be leveraged for
predicting the presence of absence of functional annotations, for
example, by using a logistic regression.  However, there is not enough
signal in the data to independently predict annotations in other
functional categories.  The empirical Bayes estimates of the
hyper-parameters that support these conclusions in the various types
of analyses are consistent; $\hat \alpha<1$ and small; and $\hat B$
nearly block diagonal with two positive blocks comprising the four
identifiable protein complexes.  In these previous analyses, we fixed the
number of latent protein complexes to 15; the number of broad functional categories in Table \ref{tab:functions}.

The latent protein complexes are not a-priori identifiable in our
model.  To resolve this, we estimated a mapping between latent complexes and
functions by minimizing the divergence between true and predicted
marginal frequencies of membership, where the truth was evaluated on a
small fraction of the interactions.  We used this mapping to compare
predicted versus known functional annotations for all proteins.  The
best estimated mapping is shown in the left panel of Figure \ref{fig:identify}, along with the marginal latent category membership, and it is compared to the 15 broad functional categories Table \ref{tab:functions}, along with the known category membership (in the MIPS database), in the right panel.
 Figure \ref{fig:exampleprot} displays a few examples of predicted
mixed membership probabilities against the true annotations, given the
\textit{estimated mapping} of latent protein complexes to functional
categories.

\subsection{Measuring the Functional Content in the Posterior}
\label{sec:functional_content}

 In a follow-up study we considered the gene ontology (GO) \citep{Ashb:etal:2000} as the source of functional annotations to consider as ground truth in our analyses. GO is a broader and finer grained functional annotation scheme if compared to that produced by the Munich Institute for Protein Sequencing.
 Furthermore, we explored a much larger model space than in the previous study, in order to tests to what extent ALB can reduce the dimensionality of the data while revealing substantive information about the functionality of proteins that can be used to inform subsequent analyses.
 We fit models with a number blocks up to $K=225$. Thanks to our nested variational inference algorithm, we were able to perform five-fold cross-validation for each value of $K$. We determined that a fairly parsimonious model $(K^*=50)$ provides a good description of the observed protein interaction network. This fact is (qualitatively) consistent with the quality of the predictions that were obtained with a parsimonious model ($K=15$) in the previous section, in a different setting. This finding supports the hypothesis that groups of interacting proteins in the MIPS data set encode biological signal at a scale of aggregation that is higher than that of protein complexes.\footnote{It has been recently suggested that stable protein complexes average five proteins in size \citep{Krog:Cagn:Yu:Zhon:2006}. Thus, if ALB captured biological signal at the protein-complex resolution, we would expect the optimal number of groups to be much higher (Disregarding mixed membership, $871/5\approx 175$.)}
\begin{figure}[b!]
  \centering
   \includegraphics[width=16.5cm]{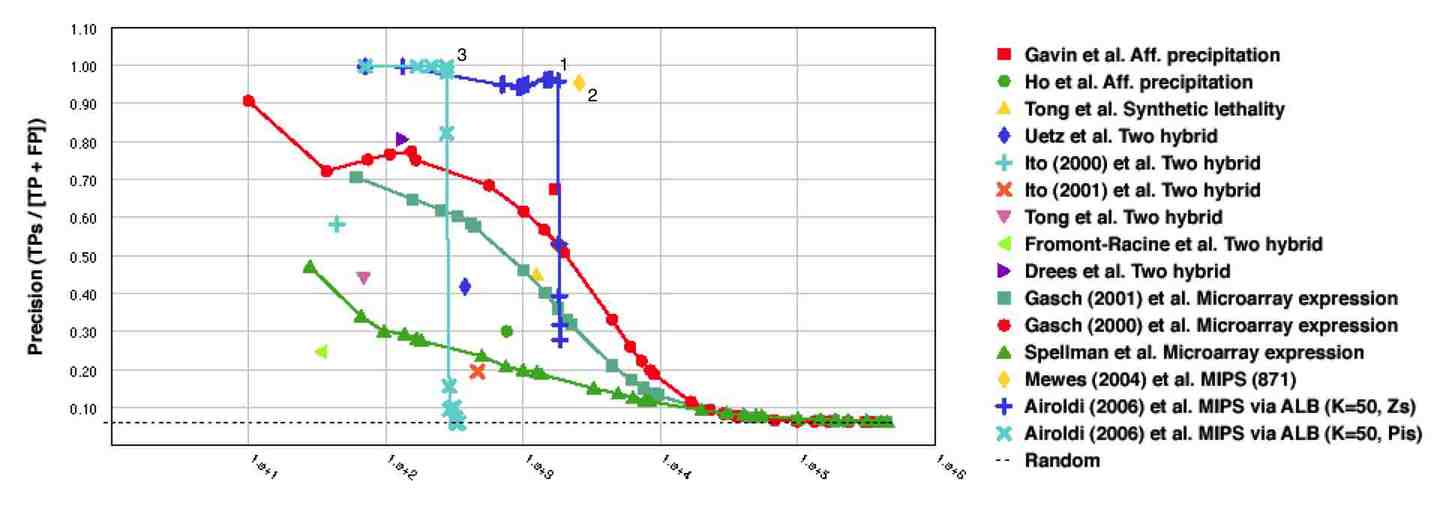}
  \caption{In the top panel we measure the functional content of the
    the MIPS collection of protein interactions (yellow diamond), and
    compare it against other published collections of interactions and
    microarray data, and to the posterior estimates of the ALB
    models---computed as described in Section
    \ref{sec:functional_content}. A breakdown of three estimated
    interaction networks (the points annotated 1, 2, and 3) into most represented
    gene ontology categories is detailed in Table
    \ref{tab:detail_functional_content}.}
\label{fig:functional_content}
\end{figure}

 We settled on a model with $K^*=50$ blocks.
 To evaluate the functional content of the interactions predicted by such model, we first computed the posterior probabilities of interactions by thresholding the posterior expectations
\[
\mathbb{E} \bigm[ R(p,q)=1 \bigm] \approx \widehat{\vec{\pi}}_p\,' \,
\widehat{B} ~ \, \widehat{\vec{\pi}}_q \qquad \hbox{ and } \qquad
\mathbb{E} \bigm[ R(p,q)=1 \bigm] \approx \widehat{\vec{\phi}}_{\ptoq}\,' \,
\widehat{B} ~ \, \widehat{\vec{\phi}}_{\pfromq},
\]
 and we then computed the precision-recall curves corresponding to these predictions. These curves are shown in Figure \ref{fig:functional_content} as the light blue ($-\times$) line and the the dark blue ($-+$) line.
 In Figure \ref{fig:functional_content} we also plotted the functional content of the original MIPS collection.
\begin{table}[t!]
\begin{center}
\begin{tabular}{rlp{9cm}rr} \hline
\# & GO Term    & Description                                   & Pred. & Tot.   \\ \hline
 1 & GO:0043285 & Biopolymer catabolism                         &   561 &  17020 \\
 1 & GO:0006366 & Transcription from RNA polymerase II promoter &   341 &  36046 \\
 1 & GO:0006412 & Protein biosynthesis                          &   281 & 299925 \\
 1 & GO:0006260 & DNA replication                               &   196 &   5253 \\
 1 & GO:0006461 & Protein complex assembly                      &   191 &  11175 \\
 1 & GO:0016568 & Chromatin modification                        &   172 &  15400 \\
 1 & GO:0006473 & Protein amino acid acetylation                &    91 &    666 \\
 1 & GO:0006360 & Transcription from RNA polymerase I promoter  &    78 &    378 \\
 1 & GO:0042592 & Homeostasis                                   &    78 &   5778 \\ \hline
 2 & GO:0043285 & Biopolymer catabolism                         &   631 &  17020 \\
 2 & GO:0006366 & Transcription from RNA polymerase II promoter &   414 &  36046 \\
 2 & GO:0016568 & Chromatin modification                        &   229 &  15400 \\
 2 & GO:0006260 & DNA replication                               &   226 &   5253 \\
 2 & GO:0006412 & Protein biosynthesis                          &   225 & 299925 \\
 2 & GO:0045045 & Secretory pathway                             &   151 &  18915 \\
 2 & GO:0006793 & Phosphorus metabolism                         &   134 &  17391 \\
 2 & GO:0048193 & Golgi vesicle transport                       &   128 &   9180 \\
 2 & GO:0006352 & Transcription initiation                      &   121 &   1540 \\ \hline
 3 & GO:0006412 & Protein biosynthesis                          &   277 & 299925 \\
 3 & GO:0006461 & Protein complex assembly                      &   190 &  11175 \\
 3 & GO:0009889 & Regulation of biosynthesis                    &    28 &    990 \\
 3 & GO:0051246 & Regulation of protein metabolism              &    28 &    903 \\
 3 & GO:0007046 & Ribosome biogenesis                           &    10 &  21528 \\
 3 & GO:0006512 & Ubiquitin cycle                               &     3 &   2211 \\ \hline
\end{tabular}
\end{center}
\caption{Breakdown of three example interaction networks into most represented gene ontology categories---see text for more details. The digit in the first column indicates the example network in Figure \ref{fig:functional_content} that any given line refers to. The last two columns quote the number of predicted, and possible pairs for each GO term.}
\label{tab:detail_functional_content}
\end{table}
 This plot confirms that the MIPS collection of interactions, our data, is one of the most precise (the $Y$ axis measures precision) and most extensive (the $X$ axis measures the amount of functional annotations predicted, a measure of recall) source of biologically relevant interactions available to date---the yellow diamond, point \# 2.
\begin{sidewaysfigure}
  \centering
   \includegraphics[width=21.5cm]{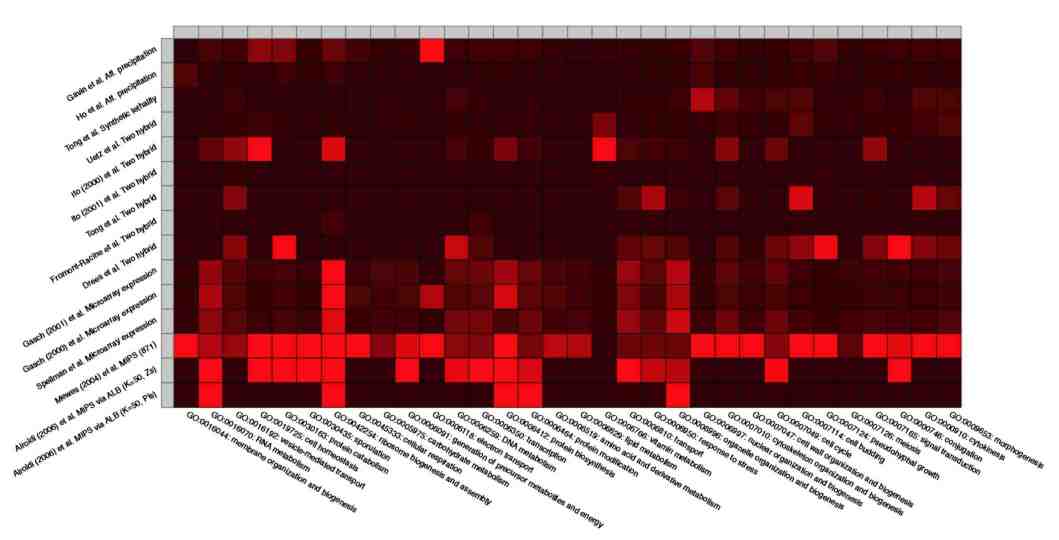}
  \caption[]{We investigate the correlations between data collections
    (rows) and a sample of gene ontology categories (columns). The
    intensity of the square (red is high) measures the area under the
    precision-recall curve.}
\label{fig:GO_correlations}
\end{sidewaysfigure}
 The posterior means of $(\vec{\pi}_{1:N})$ and the estimates of $(\alpha,B)$ provide a parsimonious representation for the MIPS collection, and lead to precise interaction estimates, in moderate amount (the light blue, $-\times$ line).  The posterior means of $(Z_\rightarrow,Z_\leftarrow)$ provide a richer representation for the data, and describe most of the functional content of the MIPS collection with high precision (the dark blue, $-+$ line).
 Most importantly, notice the estimated protein interaction networks, i.e., pluses and crosses, corresponding to lower levels of recall feature a more precise functional content than the original. This means that the proposed latent block structure is helpful in summarizing the collection of interactions---by ranking them properly. (It also happens that dense blocks of predicted interactions contain known functional predictions that were not in the MIPS collection.)
 Table \ref{tab:detail_functional_content} provides more information about three instances of predicted interaction networks displayed in Figure \ref{fig:functional_content}; namely, those corresponding the points annotated with the numbers 1 (a collection of interactions predicted with the $\vec{\pi}$'s), 2 (the original MIPS collection of interactions), and 3 (a collection of interactions predicted with the $\vec{\phi}$'s).
 Specifically, the table shows a breakdown of the predicted (posterior) collections of interactions in each example network into the gene ontology categories. A count in the second-to-last column of Table \ref{tab:detail_functional_content} corresponds to the fact that both proteins are annotated with the same GO functional category.\footnote{Note that, in GO, proteins are typically annotated to multiple functional categories.} 
 Figure \ref{fig:GO_correlations} investigates the correlations between the data sets (in rows) we considered in Figure \ref{fig:functional_content} and few gene ontology categories (in columns).  The intensity of the square (red is high) measures the area under the precision-recall curve \citep{Myer:Barr:Hibb:Hutt:2006}.

 In this application, the ALB learned information about (i) the mixed membership of objects to latent groups, and (ii) the connectivity patterns among latent groups.  These quantities were useful in describing and summarizing the functional content of the MIPS collection of protein interactions. This suggests the use of ALB as a dimensionality reduction approach that may be useful for performing model-driven de-noising of new collections of interactions, such as those measured via high-throughput experiments.

\section{Conclusions}

When applied to a sample of measurements on pairs of objects,
\textit{Admixture of Latent Blocks} simultaneously extracts
information about (i) the mixed membership of objects to latent
aspects, and (ii) the connectivity patterns among latent aspects,
using a nested variational EM algorithm.

We found it useful in describing and summarizing the functional
content of a protein interaction network, and we envision its use for
de-noising new collection of interactions from high-throughput
experiments.

\vskip 0.2in

\subsection*{Acknowledgments}
 This work was partially supported by National Institutes of Health under Grant No. R01 AG023141-01, by the Office of Naval Research under Contract No. N00014-02-1-0973, by the National Science Foundation under Grant No. DMS-0240019, Grant No. IIS0218466, and Grant No. DBI-0546594, and the Department of Defense, all to Carnegie Mellon University. Additional support was provided by the center for Computational Analysis of Social and Organizational Systems (CASOS) at Carnegie Mellon University.

\bibliographystyle{plainnat}

\begin{thebibliography}{28}
\providecommand{\natexlab}[1]{#1}
\providecommand{\url}[1]{\texttt{#1}}
\expandafter\ifx\csname urlstyle\endcsname\relax
  \providecommand{\doi}[1]{doi: #1}\else
  \providecommand{\doi}{doi: \begingroup \urlstyle{rm}\Url}\fi

\bibitem[Airoldi et~al.(2005)Airoldi, Blei, Xing, and
  Fienberg]{Airo:Blei:Xing:Fien:2005}
E.~M. Airoldi, D.~M. Blei, E.~P. Xing, and S.~E. Fienberg.
\newblock A latent mixed-membership model for relational data.
\newblock In \emph{ACM SIGKDD Workshop on Link Discovery: Issues, Approaches
  and Applications}, 2005.

\bibitem[Airoldi et~al.(2007)Airoldi, Blei, Fienberg, and
  Xing]{Airo:Blei:Fien:Xing:2006b}
E.~M. Airoldi, D.~M. Blei, S.~E. Fienberg, and E.~P. Xing.
\newblock Mixed membership stochastic blockmodels.
\newblock URL \url{http://arxiv.org/abs/0705.4485}.
\newblock Manuscript under review, 2007.

\bibitem[Alberts et~al.(2002)Alberts, Johnson, Lewis, Raff, Roberts, and
  Walter]{Albe:John:Lewi:Raff:2002}
B.~Alberts, A.~Johnson, J.~Lewis, M.~Raff, K.~Roberts, and P.~Walter.
\newblock \emph{Molecular Biology of the Cell}.
\newblock Garland, 4th edition, 2002.

\bibitem[Ashburner et~al.(2000)Ashburner, Ball, Blake, Botstein, Butler,
  Cherry, Davis, Dolinski, Dwight, Eppig, Harris, Hill, {Issel-Tarver},
  Kasarskis, Lewis, Matese, Richardson, Ringwald, Rubinand, and
  Sherlock]{Ashb:etal:2000}
M.~Ashburner, C.~A. Ball, J.~A. Blake, D.~Botstein, H.~Butler, J.~M. Cherry,
  A.~P. Davis, K.~Dolinski, S.~S. Dwight, J.~T. Eppig, M.~A. Harris, D.~P.
  Hill, L.~{Issel-Tarver}, A.~Kasarskis, S.~Lewis, J.~C. Matese, J.~E.
  Richardson, M.~Ringwald, G.~M. Rubinand, and G.~Sherlock.
\newblock Gene ontology: {T}ool for the unification of biology. {T}he gene
  ontology consortium.
\newblock \emph{Nature Genetics}, 25\penalty0 (1):\penalty0 25--29, 2000.

\bibitem[Barnard et~al.(2003)Barnard, Duygulu, de~Freitas, Forsyth, Blei, and
  Jordan]{Barn;Duyg;Frei;Fors;2003}
K.~Barnard, P.~Duygulu, N.~de~Freitas, D.~Forsyth, D.~Blei, and M.~Jordan.
\newblock Matching words and pictures.
\newblock \emph{Journal of Machine Learning Research}, 3:\penalty0 1107--1135,
  2003.

\bibitem[Blei et~al.(2003)Blei, Jordan, and Ng]{Blei:Jord:Ng:2003}
D.~M. Blei, M.~I. Jordan, and A.~Y. Ng.
\newblock Hierarchical bayesian models for applications in information
  retrieval.
\newblock In J.~M. Bernardo, M.~J. Bayarri, J.~O. Berger, A.~P. Dawid,
  D.~Heckerman, A.~F.~M. Smith, and M.~West, editors, \emph{Bayesian
  Statistics}, volume~7, pages 25--44. Oxford University Press, 2003.

\bibitem[Carlin and Louis(2005)]{Carl:Loui:2005}
B.~P. Carlin and T.~A. Louis.
\newblock \emph{Bayes and {E}mpirical {B}ayes Methods for Data Analysis}.
\newblock Chapman \& Hall, second edition, 2005.

\bibitem[Cohn and Hofmann(2001)]{Cohn:Hofm:2001}
D.~Cohn and T.~Hofmann.
\newblock The missing link---{A} probabilistic model of document content and
  hypertext connectivity.
\newblock In \emph{Advances in Neural Information Processing Systems 13}, 2001.

\bibitem[Erosheva and Fienberg(2005)]{Eros:Fien:2005}
E.~A. Erosheva and S.~E. Fienberg.
\newblock Bayesian mixed membership models for soft clustering and
  classification.
\newblock In C.~Weihs and W.~Gaul, editors, \emph{Classification---The
  Ubiquitous Challenge}, pages 11--26. Springer-Verlag, 2005.

\bibitem[Erosheva et~al.(2004)Erosheva, Fienberg, and
  Lafferty]{Eros:Fien:Laff:2004}
E.~A. Erosheva, S.~E. Fienberg, and J.~Lafferty.
\newblock Mixed-membership models of scientific publications.
\newblock \emph{Proceedings of the National Academy of Sciences}, 97\penalty0
  (22):\penalty0 11885--11892, 2004.

\bibitem[Fienberg et~al.(1985)Fienberg, Meyer, and
  Wasserman]{Fien:Meye:Wass:1985}
S.~E. Fienberg, M.~M. Meyer, and S.~Wasserman.
\newblock Statistical analysis of multiple sociometric relations.
\newblock \emph{Journal of the American Statistical Association}, 80:\penalty0
  51--67, 1985.

\bibitem[Gavin et~al.(2002)Gavin, Bosche, Krause, Grandi, Marzioch, Bauer,
  Schultz, and {et. al.}]{Gavi:Bosc:Krau:etal:2002}
A.~C. Gavin, M.~Bosche, R.~Krause, P.~Grandi, M.~Marzioch, A.~Bauer,
  J.~Schultz, and {et. al.}
\newblock Functional organization of the yeast proteome by systematic analysis
  of protein complexes.
\newblock \emph{Nature}, 415:\penalty0 141--147, 2002.

\bibitem[Ho et~al.(2002)Ho, Gruhler, Heilbut, Bader, Moore, Adams, Millar,
  Taylor, Bennett, and {et. al}]{Ho:Gruh:Heil:etal:2002}
Y.~Ho, A.~Gruhler, A.~Heilbut, G.~D. Bader, L.~Moore, S.~L. Adams, A.~Millar,
  P.~Taylor, K.~Bennett, and K.~Boutilier {et. al}.
\newblock Systematic identification of protein complexes in saccharomyces
  cerevisiae by mass spectrometry.
\newblock \emph{Nature}, 415:\penalty0 180--183, 2002.

\bibitem[Hofmann(1999)]{Hofm:1999}
T.~Hofmann.
\newblock Probabilistic latent semantic analysis.
\newblock In \emph{Uncertainty in Artificial Intelligence}, volume~15, 1999.

\bibitem[Holland and Leinhardt(1975)]{Holl:Lein:1975}
P.~W. Holland and S.~Leinhardt.
\newblock Local structure in social networks.
\newblock In D.~Heise, editor, \emph{Sociological Methodology}, pages 1--45.
  Jossey-Bass, 1975.

\bibitem[Kemp et~al.(2004)Kemp, Griffiths, and Tenenbaum]{Kemp:Grif:Tene:2004}
C.~Kemp, T.~L. Griffiths, and J.~B. Tenenbaum.
\newblock Discovering latent classes in relational data.
\newblock Technical Report AI Memo 2004-019, MIT, 2004.

\bibitem[Kemp et~al.(2006)Kemp, Tenenbaum, Griffiths, Yamada, and
  Ueda]{Kemp:Tene:Grif:etal:2006}
C.~Kemp, J.~B. Tenenbaum, T.~L. Griffiths, T.~Yamada, and N.~Ueda.
\newblock Learning systems of concepts with an infinite relational model.
\newblock In \emph{Proceedings of the 21st National Conference on Artificial
  Intelligence}, 2006.

\bibitem[Krogan et~al.(2006)Krogan, Cagney, Yu, Zhong, Guo, Ignatchenko, Li,
  Pu, Datta, Tikuisis, Punna, Peregrin-Alvarez, Shales, Zhang, Davey, Robinson,
  Paccanaro, Bray, Sheung, Beattie, Richards, Canadien, Lalev, Mena, Wong,
  Starostine, Canete, Vlasblom, Wu, Orsi, Collins, Chandran, Haw, Rilstone,
  Gandi, Thompson, Musso, {St Onge}, Ghanny, Lam, Butland, Altaf-Ul, Kanaya,
  Shilatifard, O'Shea, Weissman, Ingles, Hughes, Parkinson, Gerstein, Wodak,
  Emili, and Greenblatt]{Krog:Cagn:Yu:Zhon:2006}
N.~J. Krogan, G.~Cagney, H.~Yu, G.~Zhong, X.~Guo, A.~Ignatchenko, J.~Li, S.~Pu,
  N.~Datta, A.~P. Tikuisis, T.~Punna, J.~M. Peregrin-Alvarez, M.~Shales,
  X.~Zhang, M.~Davey, M.~D. Robinson, A.~Paccanaro, J.~E. Bray, A.~Sheung,
  B.~Beattie, D.~P. Richards, V.~Canadien, A.~Lalev, F.~Mena, P.~Wong,
  A.~Starostine, M.~M. Canete, J.~Vlasblom, S.~Wu, C.~Orsi, S.~R. Collins,
  S.~Chandran, R.~Haw, J.~J. Rilstone, K.~Gandi, N.~J. Thompson, G.~Musso,
  P.~{St Onge}, S.~Ghanny, M.~H.~Y. Lam, G.~Butland, A.~M. Altaf-Ul, S.~Kanaya,
  A.~Shilatifard, E.~O'Shea, J.~S. Weissman, C.~J. Ingles, T.~R. Hughes,
  J.~Parkinson, M.~Gerstein, S.~J. Wodak, A.~Emili, and J.~F. Greenblatt.
\newblock Global landscape of protein complexes in the yeast {S}accharomyces
  {C}erevisiae.
\newblock \emph{Nature}, 440\penalty0 (7084):\penalty0 637--643, 2006.

\bibitem[Manton et~al.(1994)Manton, Woodbury, and Tolley]{Mant:Wood:Toll:1994}
K.~G. Manton, M.~A. Woodbury, and H.~D. Tolley.
\newblock \emph{Statistical Applications Using Fuzzy Sets}.
\newblock Wiley, 1994.

\bibitem[Mewes et~al.(2004)Mewes, Amid, Arnold, Frishman, Guldener, and {et.
  al}]{Mewe:Amid:Arno:etal:2004}
H.~W. Mewes, C.~Amid, R.~Arnold, D.~Frishman, U.~Guldener, and {et. al}.
\newblock Mips: analysis and annotation of proteins from whole genomes.
\newblock \emph{Nucleic Acids Research}, 32:\penalty0 D41--44, 2004.

\bibitem[Morris(1983)]{Morr:1983}
C.~Morris.
\newblock Parametric empirical {B}ayes inference: {T}heory and applications.
\newblock \emph{Journal of the American Statistical Association}, 78\penalty0
  (381):\penalty0 47--65, 1983.
\newblock With discussion.

\bibitem[Myers et~al.(2006)Myers, Barret, Hibbs, Huttenhower, and
  Troyanskaya]{Myer:Barr:Hibb:Hutt:2006}
C.~L. Myers, D.~A. Barret, M.~A. Hibbs, C.~Huttenhower, and O.~G. Troyanskaya.
\newblock Finding function: {A}n evaluation framework for functional genomics.
\newblock \emph{BMC Genomics}, 7\penalty0 (187), 2006.

\bibitem[Pritchard et~al.(2000)Pritchard, Stephens, and
  Donnelly]{Prit:Step:Donn:2000}
J.~Pritchard, M.~Stephens, and P.~Donnelly.
\newblock Inference of population structure using multilocus genotype data.
\newblock \emph{Genetics}, 155:\penalty0 945--959, 2000.

\bibitem[Rosenberg et~al.(2002)Rosenberg, Pritchard, Weber, Cann, Kidd,
  Zhivotovsky, and Feldman]{Rose:Prit:Webe:etal:2002}
N.~A. Rosenberg, J.~K. Pritchard, J.~L. Weber, H.~M. Cann, K.~K. Kidd, L.~A.
  Zhivotovsky, and M.~W. Feldman.
\newblock Genetic structure of human populations.
\newblock \emph{Science}, 298:\penalty0 2381--2385, 2002.

\bibitem[Snijders(2002)]{Snij:2002}
T.~A.~B. Snijders.
\newblock Markov chain monte carlo estimation of exponential random graph
  models.
\newblock \emph{Journal of Social Structure}, 2002.

\bibitem[Taskar et~al.(2003)Taskar, Wong, Abbeel, and
  Koller]{Task:Wong:Abbe:Koll:2003}
B.~Taskar, M.~F. Wong, P.~Abbeel, and D.~Koller.
\newblock Link prediction in relational data.
\newblock In \emph{Neural Information Processing Systems 15}, 2003.

\bibitem[Wasserman and Pattison(1996)]{Wass:Patt:1996}
S.~Wasserman and P.~Pattison.
\newblock Logit models and logistic regression for social networks: I. an
  introduction to markov graphs and $p^{\ast}$.
\newblock \emph{Psychometrika}, 61:\penalty0 401--425, 1996.

\bibitem[Xing et~al.(2003)Xing, Ng, Jordan, and Russel]{Xing:Ng:Jord:Russ:2003}
E.~P. Xing, A.~Y. Ng, M.~I. Jordan, and S.~Russel.
\newblock Distance metric learning with applications to clustering with side
  information.
\newblock In \emph{Advances in Neural Information Processing Systems},
  volume~16, 2003.

\end{thebibliography}

\end{document}